\documentclass[preprint,showpacs,superscriptaddress,nofootinbib]{revtex4}
\pdfoutput=1
\usepackage{amsfonts,amsmath,amssymb}
\usepackage{enumerate}
\usepackage{hyperref}
\usepackage{graphicx}
\usepackage{slashed}

\newcommand{\be}{\begin{equation}}
\newcommand{\ee}{\end{equation}}

\newcommand{\bea}{\begin{eqnarray}}
\newcommand{\eea}{\end{eqnarray}}

\newcommand{\bes}{\begin{subequations}}
\newcommand{\ees}{\end{subequations}}

\newcommand{\nn}{\nonumber}

\newcommand{\ra}{\rightarrow}

\newcommand{\z}{\zeta}
\newcommand{\e}{\epsilon}
\newcommand{\ep}{\epsilon}
\newcommand{\G}{\Gamma}
\newcommand{\La}{\Lambda}

\begin{document}

\title{One-loop corrections to holographic Wilson loop in $AdS_4\times \mathbb{CP}^3$}
%\preprint{\hepth{yymm.nnnn}}
\date{\today}

\author{Hyojoong Kim}
%\email{h.kim@khu.ac.kr}
\affiliation{Department of Physics
and Research Institute of Basic Science, \\ Kyung Hee University, 
 Hoegi-dong, Dongdaemun-gu,
 Seoul 130-701, Korea}
\author{Nakwoo Kim}
\email{nkim@khu.ac.kr}
\affiliation{Department of Physics
and Research Institute of Basic Science, \\ Kyung Hee University, 
 Hoegi-dong, Dongdaemun-gu,
 Seoul 130-701, Korea}
\affiliation{School of Natural Sciences, Institute for Advanced Study, \\ Princeton, NJ 08540, USA}
\author{Jung Hun Lee}
\affiliation{Department of Physics
and Research Institute of Basic Science, \\ Kyung Hee University, 
 Hoegi-dong, Dongdaemun-gu,
 Seoul 130-701, Korea}

\begin{abstract}
The evaluation  of BPS Wilson loops in ${\cal N}=6, D=3$ Chern-Simons matter theory is reduced to ordinary matrix integrals via localization technique. It is easy to check that the vacuum expectation value of 1/2 BPS Wilson loops at leading order in planar limit agrees with the regularized classical string action, via AdS/CFT. Then the subleading terms in principle can be calculated by treating the string theory semi-classically. In this article we calculate the one-loop determinant for the fluctuation modes of  holographic Wilson loop as IIA string in the dual geometry $AdS_4\times \mathbb{CP}^3$. The fermionic normal mode frequencies are expressed in terms of the hypergeometric function, and we compute the one-loop effective action numerically. The discrepancy with localization formula is due to the zero mode normalization constant, which is yet to be determined.

\end{abstract}
%\keywords{M-theory, M(atrix) Theories}
%\thispagestyle{empty}
\maketitle

\pagestyle{plain}
%%%%%%%%%%%%%%%%%%%%%%%%%%%%%%%%%%%%%%%%%
\section{Introduction}
Wilson loops are essential objects in the study of gauge field theories. In the context of the AdS/CFT correspondence \cite{Maldacena:1997re}, they have dual description as a macroscopic fundamental string  \cite{Rey:1998ik,Maldacena:1998im}. In this article, we are mainly interested in the M2-brane conformal field theory as Chern-Simons matter model, suggested by Aharony, Bergman, Jafferis and Maldacena (ABJM) \cite{Aharony:2008ug}. The supersymmetric Wilson loop operators in ABJM model, with dual geometry $AdS_4\times \mathbb{CP}^3$, are studied earlier in \cite{Drukker:2008zx,Chen:2008bp,Rey:2008bh,Drukker:2009hy}. 
%In particular the elusive 1/2-BPS Wilson loops were constructed using a supermatrix model in \cite{Drukker:2009hy}. 
The  computation of their expectation values can be greatly simplified if one utilizes the localization technique \cite{Pestun:2007rz}: when we put the gauge theory on $S^3$, the full path integral is reduced to an ordinary matrix integral \cite{Kapustin:2009kz}. It is a fascinating achievement that at strong coupling the free energy 
scales as $N^{3/2}$ and the coefficient is related to the internal space $S^7$, precisely as predicted by AdS/CFT \cite{Marino:2009jd,Drukker:2010nc}.

According to the matrix model calculation at strong coupling and planar limit, the 1/2-BPS circular Wilson loop's vacuum expectation value is (up to a framing-dependent phase)
\be
\langle W \rangle \approx \frac{1}{2} e^{\sqrt{2\pi^2(\lambda-1/24)}} ,
% \quad \mbox{i.e.}\;
%\ln \langle W\rangle = \sqrt{2\pi^2\lambda} - \ln 2 - \frac{\sqrt{2}\pi}{48}\frac{1}{\sqrt{\lambda}} + \ldots
\ee
where $\lambda$ is the 't Hooft coupling constant. On the other hand, the gravity side computation from classical string solution is $e^{\sqrt{2\pi^2\lambda}}$. The next-order correction for $S\equiv-\ln \langle W\rangle$ should be $\ln 2\approx 0.69$, and it is our goal to see if this number can be reproduced as one-loop correction on string world sheet. 

From the fluctuation lagrangian around 1/2-BPS holographic circular Wilson loop,  we find that 
the string one-loop determinant  is given as
\bea
e^{-\Gamma}=\frac{\det^{}(-\nabla^2_F-\frac{1}{2})\det^{3}(-\nabla^2_F+\frac{1}{2})}{\det^{}(-\nabla^2+2)\det^{3}(-\nabla^2)} \, . 
\label{fd}
\eea
It turns out that part of the fermionic normal mode frequencies in the numerator are given in terms of hypergeometric functions. 
This is in contrast with the Wilson loop of IIB string in $AdS_5\times S^5$, where the frequencies are logarithm of rational functions and the sum is given exactly using the Gamma function \cite{Kruczenski:2008zk}. 
We evaluate $\G$ numerically and extract the finite piece after regularization, and obtain $\Gamma_{reg} \approx -1.1 $. 

In Section \ref{flag} we setup the notation and calculate the quadratic lagrangian for string fluctuation around 1/2-BPS circular Wilson loop. In Section \ref{dreg}, we calculate the normal modes and discuss how their sum can be regularized numerically. In Section \ref{discuss} we discuss how to resolve the discrepancy between field theory and supergravity side results.
%In Section \ref{discussion} we give a summary and discussion. 
%In Appendix one can find some properties of the hypergeometric function. 

%%%%%%%%%%%%%%%%%%%%%%%%
%%%%%%%%%%%%%%%%%%%%%%%%
\section{Open strings and their fluctuation lagrangian}
\label{flag}
%%%%%%%%%%%%%%%%%%%%%%%%%%%
%%%%%%%%%%%%%%%%%%%%%%%%%%%%%%%%%%%%
We consider type IIA open strings in $AdS_4\times \mathbb{CP}^3$ background which preserve $\tfrac{3}{4}$ supersymmetry. This geometry is conjectured to be dual to ${\cal N}=6, D=3$ Chern-Simons field theory \cite{Aharony:2008ug} with $U(N)\times U(N)$ gauge symmetry and levels $(k,-k)$. In the convention we adopt here, the $D=10$ supergravity solution takes the following form.
\bea
ds^2&=& R^2_s(ds^2_{AdS_4}+4ds^2_{\mathbb{CP}^3}),\,\,\,\, e^{2\phi}=\frac{R^2_s}{k^2} , \nn\\ 
F_2&=& k{J}_{\mathbb{CP}^3},\,\,\,\, F_4=\frac{3k R^2_s}{8} \, \text{Vol}_{AdS_4} \, .
\eea
$R_s$ sets the length scale of this background, the metric tensor $ds^2_{AdS_4},ds^2_{\mathbb{CP}^3}$ are scaled to have radius one, and $J_{\mathbb{CP}^3}$ represents the K\"ahler 2-form of the internal space. 
%According to \cite{Aharony:2008ug,Rey:2008bh}, via 
The AdS/CFT correspondence relates the string and Chern-Simons description in the following way. 
\be
R_s/\sqrt{\alpha'} = (2\pi^2\lambda)^{1/4} \, ,
\ee
where $\lambda\equiv N/k$ is the 't Hooft coupling constant. For simplicity we will henceforth set $k=1$.
% since it can be easily reinstated later. 

It is convenient for us to use Poincare coordinates for AdS space, 
\bea
ds^2_{AdS_4}=\frac{1}{z^2}(-dt^2+dr^2+r^2d\phi^2+dz^2)\,  .
\eea
%And we fix our gauge by setting the world sheet coordinates $(\tau,\sigma)=(\phi,r)$. %\nk{what is the range of $\sigma$?}
%The string Nambu-Goto action is \nk{You should give the definition of $\lambda$.} \nk{Typo corrected}
%\bea
%S=\frac{\sqrt{\lambda}}{2\pi}\int d\sigma^2\sqrt{G_{00}(X)G_{11}(X)-G_{01}^2(X)} . 
%\eea
%where $G_{mn}(X)\equiv\partial_m X^{\mu}\partial_n X^{\nu}g_{\mu\nu}(X)$.
For a circular Wilson loop with radius $1$, a simple solution is given as
\bea
z=\sqrt{1-r^2} \, .
\eea
%Note that the induced metric on the worldsheet is 
%\bea
%ds^2_{ind}=h_{00}d\tau^2+h_{11}d\sigma^2,\,\,\,\, h_{ij}=\text{diag}(\frac{r^2}{\omega^2},\frac{1}{\omega^4})
%\eea
%\bea
%ds^2_{cl}=\frac{r^2}{1-r^2} \left[\frac{dr^2}{r^2(1-r^2)}+d\phi^2\right],\,\,\,\,\,\, R^{(2)}=-2.
%\eel
In conformal gauge $r=1/\cosh\sigma$ and the induced metric on worldsheet is 
% $\frac{dr}{r\sqrt{1-r^2}}\rightarrow d\sigma$. 
\bea
ds^2_{ws}=\frac{1}{\sinh^2\sigma}(d\sigma^2+d\tau^2),\quad 0\leq\sigma<\infty,\,\,0\leq\tau <2\pi \, .
\label{ws}
\eea
Note that this is hyperbolic with scalar curvature $R^{(2)}=-2$.
In order to regularize the divergence of the classical action, 
we introduce a cutoff at $z=\epsilon$ or equivalently at $\sigma=\epsilon_0$ which are related via $\epsilon=\tanh\epsilon_0$. The regularized value of the classical action is \cite{Drukker:2008zx,Chen:2008bp,Rey:2008bh}
\be
S_0 = -R^2_s/\alpha' = - \sqrt{2\pi^2 \lambda} . 
\ee
%\bea
%\omega^{01}_\tau=-\frac{\cosh\sigma}{\sinh\sigma},\,\,\,\,\,\,\, 
%\nabla_0=\partial_\tau-\frac{1}{2}\frac{\cosh\sigma}{\sinh\sigma}\Gamma_{01},\,\,\,\,\,\,\, 
%\nabla_1=\partial_\sigma                \label{sc}
%\eea

Now we are to consider the fluctuation modes around this classical solution. Similar computations have been performed in a number of articles including \cite{Kruczenski:2008zk,McLoughlin:2008ms,Alday:2008ut,Krishnan:2008zs,Bandres:2009kw,
Beccaria:2012qd,Forini:2012bb}. 
For the bosonic sector, the computations should be very similar to those of the circular Wilson loop in $AdS_5\times S^5$ presented in \cite{Kruczenski:2008zk}. One easily finds that after the gauge fixing, there are two modes from AdS space with effective mass parameter 2, and there are six massless modes from $\mathbb{CP}^3$. Altogether they account for the denominator of \eqref{fd}.

%%%%%%%%%%%%%%%%%%%%%%%%%%%%%%%%%%%%%%%%%%%%%%%%%%%%%%%%%%%%%%%%%%%%
%\subsection{The bosonic sector}
%We get the following bosonic flucuation Lagrangian up to quadratic order
%\bea
%\mathcal{L}_{2B}^{AdS_4}=\frac{1}{2}\sqrt{h}[h^{00}(\partial_{\tau}\tilde{\eta})^2+h^{11}(\partial_{\sigma}\tilde{\eta})^2+2\tilde{\eta}^2+h^{00}(\partial_{\tau}\tilde{\omega})^2+h^{11}(\partial_{\sigma}\tilde{\omega})^2+2\tilde{\omega}^2], \label{bads}
%\eea
%where $\tilde{\eta}=\tilde{x}/\omega$.
%\bea
%\mathcal{L}_{2B}^{\mathbb{C}P^3}=\frac{1}{2}\sqrt{h}\tilde{\zeta}^a\Delta\tilde{\zeta}^a, \label{bcp}
%\eea
%where the index $a$ runs 4 to 9.

%%%%%%%%%%%%%%%%%%%%%%%%%%%%%%%%%%%%%%%%%%%%%%%%%%%%%%%%%%%%%%%%%%%%%
%\subsection{The Green-Schwarz action in the fermionic sector}
For the fermionic part, up to 
 quadratic order the $\kappa$-symmetric Green-Schwarz action is
\bea
S_F=-\frac{i R^2_s}{2\pi\alpha'}\int d^2\sigma
(\sqrt{h}h^{ab}\delta^{IJ}-\epsilon^{ab}S^{IJ})\bar{\theta}^I\rho_a D_b^{JK}\theta^K \, . 
\eea
where $S^{IJ}=\text{diag}(1,-1)$, $\rho_a=\Gamma_A\partial_aX^M E^A_M$.  $X^M$ parametrizes the ten-dimensional spacetime, $\Gamma_A$ is gamma matrix, $E^A_{M}$ is vielbein, and $h_{ab}$ is the worldsheet metric. The spinors $\theta^1$ and $\theta^2$ have opposite chirality, i.e. 
\bea
\Gamma_{11}\theta^1=\theta^1 \, ,  \quad
\Gamma_{11}\theta^2=-\theta^2\,.
\eea
The covariant derivative for spinor field is spelt out as \cite{McLoughlin:2008ms}
\bea
D_a^{JK}&=&\biggl(\partial_a+\frac{1}{4}\partial_a X^M\omega_M^{AB}\Gamma_{AB}\biggr)
-\frac{1}{8}\partial_aX^ME^A_MH_{ABC}\Gamma^{BC}(\sigma_3)^{JK} \\ \nonumber
&+&\frac{1}{8}e^{\phi}[F_{(0)}(\sigma_1)^{JK}+\slashed{F}_{(2)}(i\sigma_2)^{JK}+\slashed{F}_{(4)}(\sigma_1)^{JK}]\rho_a \, .
\eea
%where \nk{chaged, and we still need $F_{(0)}$}
%\bea
%e^{\phi}(F_{(2)})_{\mu\nu}=\frac{1}{R_{AdS}}J_{\mu\nu}\, , \quad 
%e^{\phi}(F_{(4)})_{\mu\nu\rho\sigma}=\frac{3}{R_{AdS}}\epsilon_{\mu\nu\rho\sigma} \, .
%\eea
After some calculation one can rewrite the fermion fluctuation lagrangian simply as 
\be
\mathcal{L}=i\bar{\Psi}\mathcal{K}\Psi\, ,
%\eea
%where the new Dirac operator is 
%\be
\quad 
\mathcal{K} = \sqrt{h}(\tau^i\nabla_i-i\Gamma_{3/4}\Gamma_{01}) \, . 
\ee
We note that this expression is obtained after rotating the spinor by a 
unitary matrix 
\bea
S=\exp\biggl(\frac{\alpha}{2}\Gamma_{13}\biggr)\, , \quad \tan\alpha=\frac{r}{z} \, . 
\eea  
$\Psi$ also satisfies
$P_+\Psi=\Psi$ with $P_+=(1+\Gamma_{01}\Gamma_{11})/2$,
$d=2$ gamma matrices $\tau^i$ satisfy $\{\tau_i, \tau_j\}=2h_{ij}$, and 
\bea
\Gamma_{3/4}=\frac{1}{4i}\left(3\Gamma_{23}+(\Gamma_{45}+\Gamma_{67}+\Gamma_{89})\Gamma_{11}\Gamma_{01}\right)\, .
\eea
It is obvious that $\Gamma_{3/4}$ is hermitian and traceless. 
When diagonalized, it can be written as for instance $\mbox{diag}(1,1,1,0)\otimes \mbox{diag}(1,1,-1,-1)$. This implies that we should have 4 massless fermionic modes, and 12 modes with mass 1, on the worldsheet. We note here that this result is in agreement with similar analysis done for instance in \cite{Bandres:2009kw,Forini:2012bb}. 

For the computation of the determinant, we might as well consider the square of Dirac operator.
We consider 
%After being normalized with $\sqrt{h}$, the square of the fermionic operator is
\bea
\Delta_{F}\equiv (i\tau^i\nabla_i+\Gamma_{3/4}\Gamma_{01})^2=
-\nabla^2_F+\frac{R^{(2)}}{4}+\Gamma^2_{3/4}\, ,
\eea
%%%%%%%%%%%%%%%%%%%%%
%%%%%%%%%%%%%%%%%%%%%
%\jh{derivation of the eq.(13)}
%\bea
%(i\tau^i\nabla_i+\Gamma_{3/4}\Gamma_{01})^2=-\tau^i\nabla_i\tau^j\nabla_j+i\Gamma_{3/4}(\tau^i\nabla_i\Gamma_{01}+\Gamma_{01}\tau^j\nabla_j)+\Gamma_{3/4}^2(\Gamma_{01})^2\label{de}
%\eea
%The first term of (\ref{de})
%\bea
%-\tau^i\nabla_i\tau^j\nabla_j&=&\frac{\omega^2}{r^2}\biggl(\partial_0+\frac{1}{2\omega}\Gamma_{01}\biggr)^2+\biggl(2r\omega^2-\frac{\omega^2}{r}\biggr)\partial_1-\omega^4\partial_1^2-\frac{1}{2} \nonumber \\
%&=&-\nabla_F^2+\biggl(r\omega^2+\frac{\omega^4}{r}-\frac{\omega^2}{r}\biggr)\partial_1-\frac{1}{2}\nonumber\\
%&=&-\nabla^2_F-\frac{1}{2},
%\eea
%where in the second line the middle term is vanished via the classical solution ($\omega=\sqrt{1-r^2}$)
where $\nabla^2_F\equiv\tfrac{1}{\sqrt{g}}\nabla_i(\sqrt{g}g^{ij}\nabla_j)$ and for the solution we have here $R^{(2)}$=$-2$. %The second term of (\ref{de}) is also vanished since $\{\tau^i,\Gamma_{01}\}=0$. So we combine the above results together we obtain the right hand side in (13).
%%%%%%%%%%%%%%%%%%%%%
%%%%%%%%%%%%%%%%%%%%%
%where $\nabla_F^2=\nabla^i\nabla_i$ with spinorial covariant derivatives. 
%Since the induced metric \eqref{ws} is independent of $\tau$, the eigenmodes for $\tau$ are simply Fourier modes $e^{i \omega \tau}\,(\omega \in \mathbb{Z}+1/2)$ and one may substitute $\partial_\tau \ra i\omega$.
%In the end we have
%Now in the basis where $\Gamma_{3/4}$ is diagonalized, we have 
%\nk{first and second lines below are dubious. In the end, we should keep only the third lines. And also $r\ra\omega$}
%\bea
%\Delta_{F}- \Gamma^2_{3/4}&=&\sinh^2\sigma(-\partial^2_\sigma+\omega^2)+\omega\Gamma_{01}\sinh\sigma\cosh\sigma+\frac{\sinh^2\sigma-1}{4}.
%\\
%\Delta'_{\gamma=\pm\frac{1}{2}}&=&\sinh^2\sigma(-\partial^2_\sigma+\omega^2)+%\frac{\sinh^2\sigma}{4}+\omega\Gamma_{01}\sinh\sigma\cosh\sigma.
%\eea
%We have introduced the Fourier mode solution in
%$\tau$ ($\sim e^{ir\tau}$), where $r$ is a half-integer number, in the second line for the above two equations.
%%%%%%%%%%%%%%%%%%%%%%%%%%%%%%%%%%%%%%%%%%%%%%%%%%%%%%%%%%%%%%%%%%%%%%%%%%%%%%%%%%%%%%%%%%%%%
%%%%%%%%%%%%%%%%%%%%%%%%%%%%%%%%%%%%%%%%%%%%%%%%%%%%%%%%%%%%%%%%%%%%%%%%%%%%%%%%%%%%%%%%%%%%%

%%%%%%%%%%%%%%%%%%%%%%%%%%%%%%%%%
%\subsection{Circular Wilson Loop}
%%%%%%%%%%%%%%%%%%%%%%%%%%%%%%%%%
Our results so far can be summarized in the following expression for one-loop partition function for fluctuation modes. %\nk{probably this is not the most general form. In numerator we keep $R^{(2)}$ as undetermined, but in denominator probably we set $R^{(2)}=-2$ already? Then this should be rewritten.} 
\bea
Z=\frac{\det^{2/2}(-\nabla^2_F-\frac{1}{2})\det^{6/2}(-\nabla^2_F+\frac{1}{2})}{\det^{2/2}(-\nabla^2+2)\det^{6/2}(-\nabla^2)} . 
\label{pf1}
\eea
Note that in the denominator $\nabla^2$ is the usual scalar Laplacian, while $\nabla^2_F$ is understood to contain spin connection for spinor fields. One can repeat the same computation for a straight line which is also 1/2-BPS and we have checked the result is again given exactly as \eqref{pf1}.

%For the given metric, one can easily calculate
% $R^{(2)}=-2$ for instance. Note that in Eq.(\ref{pf1}) $\nabla^2$ in denominator is that for a scalar field, while $\nabla^2$ in numerator is
%$\gamma_a\gamma_b \nabla^a \nabla^b$ for worldsheet fermions. 

 %For our solution $\nabla^2_F$ contains $\Gamma_{01}$ whose eigenvalues are $\pm 1$. More concretely, $Z$ can be expressed as 
 %\be
 %Z = 
 %\ee

%%%%%%%%%%%
\section{Calculation of the determinant}
\label{dreg}
%%%%%%%%%%%%%
Now let us consider the evaluation of \eqref{pf1}. Thanks to the axial symmetry of the string worldsheet, we can easily perform the mode expansion for $\tau$ variable. 
%It is straightforward to calculate the normal mode frequency for operators appearing in Eq.(\ref{pf1}). The usual separation of variables technique can be applied. 
We impose periodic (anti-periodic) boundary condition for 
bosonic (fermionic) fields. Then $Z$ can be expressed using determinants of ordinary second-order differential operators. More concretely, we have for instance %\nk{Eq(16) not consistent with (19)(20)?}
\bea
\det (-\nabla^2) &=& \prod_{n\in \mathbb{Z}}\det[\sinh^2\sigma(-\partial_{\sigma}^2+n^2)]
\\
{\det}^2 (-\nabla^2_F) &=& \prod_{\nu\in \mathbb{Z}+1/2}\det[\sinh^2\sigma(-\partial^2_{\sigma}
+\nu^2+\tfrac{1}{4}\coth^2\sigma+\nu\coth\sigma)]\nonumber\\
&\times&\det[\sinh^2\sigma(-\partial^2_{\sigma}+\nu^2+\tfrac{1}{4}\coth^2\sigma-\nu\coth\sigma)].
\eea
  The conformal factor $\sinh^2\sigma$ cancel between bosonic and fermionic determinants. We define
\bea
\omega_n^{B1}&=&\ln\left[ \det(-\partial^2_\sigma+n^2+2\text{csch}^2\sigma)/C \right],\\
\omega_n^{B3}&=&\ln\left[ \det(-\partial_\sigma^2+n^2)/C \right],\\
\omega^{F1}_\nu &=& \ln\left[
\det ( -\partial_\sigma^2 + \nu^2 + \nu \coth \sigma +\tfrac{1}{4}\coth^2\sigma+\tfrac{1}{2}{\rm csch}^2\sigma)/C
\right],
\\
\omega^{F3}_\nu &=& \ln\left[
\det ( -\partial_\sigma^2 + \nu^2 + \nu \coth \sigma +\tfrac{1}{4}\coth^2\sigma-\tfrac{1}{2}{\rm csch}^2\sigma)/C
\right] , 
\eea
where we have included $C=\det(-\partial^2_\sigma)$ as an overall normalization. 
The 1-loop effective action can be written as 
\bea
{ \Gamma}\equiv-\ln Z=\sum_{n\in Z}(\omega_n^{B1}+3\omega_n^{B3})-\frac{1}{2}\sum_{\nu\in Z+1/2}(\omega_\nu^{F1}
+\omega_{-\nu}^{F1}+3\omega_\nu^{F3}+3\omega_{-\nu}^{F3}).
\eea

It turns out that each sum $\sum\omega_n$ is divergent and there is an ordering problem. This problem is of course commonplace in quantum field theory, and for the energy correction of spinning strings in $AdS_4\times \mathbb{CP}^3$ the ordering issue has been addressed in \cite{Gromov:2008fy,Bandres:2009kw,Beccaria:2012qd}. Here we 
follow the prescription in \cite{Kruczenski:2008zk,Frolov:2004bh}: one  introduces a regulator $\mu$ in the process of synchronizing the summation indices for bosonic and fermionic modes. For small $\mu$, we have 
\bea
\Gamma_{reg}&\equiv&\sum_{n\in Z}e^{-\mu|n|}(\omega_n^{B1}+3\omega_n^{B3})-\frac{1}{2}\sum_{\nu\in Z+1/2}
e^{-\mu|\nu|}(\omega_\nu^{F1}
+\omega_{-\nu}^{F1}+3\omega_\nu^{F3}+3\omega_{-\nu}^{F3}) \nn \\
&=&\frac{1}{4}\sum_{n\in Z}\biggl[e^{-\mu|n|}\biggl(4\omega_n^{B1}+12\omega_n^{B3}-\omega_{n+1/2}^{F1}
-\omega_{n-1/2}^{F1}-\omega_{-n-1/2}^{F1}-\omega_{-n+1/2}^{F1}\nonumber\\
&&-3\omega_{n+1/2}^{F3}-3\omega_{n-1/2}^{F3}-3\omega_{-n-1/2}^{F3}-3\omega_{-n+1/2}^{F3}\biggr)\nonumber\\
&+&(e^{-\mu|n|}-e^{-\mu|n+1/2|})(\omega_{n+1/2}^{F1}+\omega_{-n-1/2}^{F1}+3\omega_{n+1/2}^{F3}+3\omega_{-n-1/2}^{F3})\nonumber\\
&+&(e^{-\mu|n|}-e^{-\mu|n-1/2|})(\omega_{n-1/2}^{F1}+\omega_{-n+1/2}^{F1}+3\omega_{n-1/2}^{F3}+3\omega_{-n+1/2}^{F3})\biggr]\nn\\
&=& \sum_{n=0}^{\infty} G_n + G' + {\cal O}(\mu) 
\label{reg}
\eea
Here we have defined 
\bea
G_0&=&\frac{1}{2}(2\omega_0^{B1}+6\omega_0^{B3}-\omega_{1/2}^{F1}-\omega_{-1/2}^{F1}-3\omega_{1/2}^{F3}-3\omega_{-1/2}^{F3})
\\
G_{n}&=&\frac{1}{2}
%e^{-\mu n}
\biggl[4\omega_n^{B1}+12\omega_n^{B3}-\omega_{n+1/2}^{F1}-\omega_{n-1/2}^{F1}
-\omega_{-n-1/2}^{F1}-\omega_{-n+1/2}^{F1} , \nonumber\\
&&-3\omega_{n+1/2}^{F3}-3\omega_{n-1/2}^{F3}-3\omega_{-n-1/2}^{F3}-3\omega_{-n+1/2}^{F3}\biggr]
,\quad (n>0)\\
G'&=&\lim_{\mu\ra 0} \frac{\mu}{4}\sum_{n>0}e^{-\mu n}\biggl[\omega_{n+1/2}^{F1}+\omega_{-n-1/2}^{F1}-\omega_{n-1/2}^{F1}-
\omega_{-n+1/2}^{F1}\nonumber\\
&&+3\omega_{n+1/2}^{F3}+3\omega_{-n-1/2}^{F3}-3\omega_{n-1/2}^{F3}-3\omega_{-n+1/2}^{F3}\biggr] . 
\eea

%%%%%%%%%%%%%%%%%%%%%
\subsection{Calculation of the frequencies}
%%%%%%%%%%%%%%%%%%
To evaluate $\omega^B_n$ and $\omega^F_\nu$, following \cite{Kruczenski:2008zk,Hou:2008pg,Beccaria:2010tb,Forini:2010ek} we utilize
the Gelfand-Yaglom theorem: 
For a differential operator $\cal{O}$ with periodic boundary condition in $\sigma\in[a,b]$, the product of all eigenvalues can be alternatively obtained by solving the homogeneous differential equation 
${\cal O} \psi = 0$ with initial condition $\psi(a)=\psi_{0}(a)=0,\,\psi'(a)=\psi'_{0}(a)=1$. 
In particular,
\be
\frac{\det\mathcal{O}}{\det\mathcal{O}_0} = \frac{\psi(b)}{\psi_{0}(b)}, 
\label{diri}
\ee
where $\mathcal{O}_{0}=-\partial^2_\sigma$. For our problem originally $\sigma$ ranges in $0<\sigma<\infty$, but we will introduce both UV and IR regulators and consider instead $\epsilon_0 < \sigma < L$. Eq.\eqref{diri} will be used for non-zero modes, while for the zero-modes we take Neumann boundary conditions at $L$, and we need to use $\psi'(b)/\psi'_0(b)$ instead on the right hand side of \eqref{diri}.

For the bosonic part the operators are exactly the same as the counterpart in $AdS_5\times S^5$ of IIB string theory, and we simply import the results in \cite{Kruczenski:2008zk}. For large $L$ ($\epsilon_0$ is not necessarily small yet.), 
\bea
\exp(\omega^{B1}_n) &=& 
\left\{
\begin{array}{l}
\frac{(|n|+\coth\epsilon_0)}{2|n|(|n|+1)} e^{|n|(L-\epsilon_0)} , \quad n\neq 0 
\cr
\coth\epsilon_0 , \quad n= 0 
\end{array}
\right.
\\
%\eea
%\bea
\exp(\omega^{B3}_n) &=& 
\left\{
\begin{array}{l}
\frac{e^{|n|(L-\epsilon_0)}}{2|n|} , \quad n\neq 0 
\cr
1 , \quad n= 0 
\end{array}
\right.
\eea

Let us now turn to the fermionic modes. 
%\bea
%\omega_r^{(s)}&=&\ln\frac{\det(-\partial^2_\sigma+r^2+\frac{1}{4}+
%\frac{2}{\sinh^2\sigma}+r\coth\sigma)}{\det(-\partial^2_\sigma)}\,\,\,(r\in\mathbb{Z}+1/2)\label{ffs}\\
%\omega_r^{(t)}&=&\ln\frac{\det(-\partial^2_\sigma+r^2+\frac{1}{4}
%+r\coth\sigma)}{\det(-\partial^2_\sigma)}\,\,\,(r\in\mathbb{Z}+1/2) \label{fft}
%\eea
%If we are to use GY theorem, we need to consider the solution  of homogeneous
%differential equations. It turns out that the solutions are given as hypergeometric functions. 
For the differential operators associated with fermionic fluctuations, 
%In order to see that, let us rewrite the differential equations. For ${\cal O}_+^{(s)}\psi = 0$, 
we find it useful to introduce a new variable
\be 
\z = \coth\sigma . 
\ee
Note that for $0<\sigma<\infty$, we have $1<\z<\infty$. 
%It is also useful to know 
%\be
%\frac{d\z}{d\sigma} = 2(1-\z) . 
%\ee
We start with the equation associated with $\omega^{F1}_\nu$. We should originally consider
\be
( -\partial_\sigma^2 + \nu^2 + \nu \coth \sigma +\tfrac{1}{4}\coth^2\sigma+\tfrac{1}{2}{\rm csch}^2\sigma)
  \psi(\sigma)= 0 . 
 \label{f1}
\ee 
%For $\nu\neq \pm \frac{1}{2}$, we set 
%\be
%\psi(\sigma) = (\zeta+1)^{-\nu/2+1/4}(\zeta-1)^{\nu/2+1/4} y(\zeta) , 
%\ee
% and one easily checks that \eqref{f1} simplies  to 
% \be
%(\zeta^2-1) y''(\zeta) + (3\zeta+2\nu) y'(\zeta) = 0 . 
%\ee
%It is easy to solve this equation. 
The two linearly independent solutions can be chosen as follows (for $\nu\neq \pm \frac{1}{2}$)
\bea
u_\nu (\sigma) &=& (\zeta+1)^{-\nu/2+1/4} (\z-1)^{\nu/2+1/4} , 
\\
v_\nu (\sigma) &=& (\z+1)^{\nu/2-1/4} (\z-1)^{-\nu/2-1/4}(2\nu-\z) . 
\eea
Writing down the solution with appropriate initial condition and taking the limit $L\ra \infty$, we obtain the following result. 
%The Wronskian is easily computed as 
%\be
%u_\nu(\sigma) v'_\nu(\sigma) - u'_\nu(\sigma) v_\nu(\sigma) = 4\nu^2-1 .
%\ee
%So the solution with appropriate initial condition at $\sigma=\epsilon_0$ is given as 
%\be
%\psi(\sigma) = \frac{1}{4\nu^2-1} ( u_\nu(\epsilon_0) v_\nu(\sigma) - v_\nu(\epsilon_0) u_\nu(\sigma)) .
%\ee
%We note that for $\nu>1/2$, $v_\nu(\infty)=\infty$ and for $\nu<-1/2$, $u_\nu(\infty)=\infty$. Thus we have
%\bea
%\omega^{F1}_{\nu} &=&
%\left\{
%\begin{array}{l}
% \ln \left[ \frac{1}{4\nu^2-1} u_\nu(\epsilon_0) v_\nu (L) \right] , \quad \nu>+1/2
%\cr
%\ln \left[ \frac{-1}{4\nu^2-1} v_\nu(\epsilon_0) u_\nu (L) \right] , 
%\quad \nu<-1/2 
%\end{array}
%\right. 
%\eea
\bea
\omega^{F1}_{\nu} &=&
\left\{
\begin{array}{l}
 \ln \left[ \frac{e^{(\nu+1/2)(L-\e_0)}}{2\nu+1} \sqrt{\frac{1+\e}{2\e^2}}  \right] , \quad \nu \ge +1/2
\cr
\ln \left[ \sqrt{\frac{1+\ep}{2\ep}}  \right] , \quad \nu = -1/2 
\cr
\ln \left[ \frac{e^{-(\nu+1/2)(L-\e_0)}}{4\nu^2-1} \frac{1-2\nu \e}{\e^2}\sqrt{\frac{2\e}{1+\e}}  \right]
 , 
\quad \nu < -1/2 
\end{array}
\right. 
\label{f1f}
\eea
Here we introduced $\ep=\tanh\ep_0$ for cutoff of $z$-coordinate. $\nu=\pm \tfrac{1}{2}$ are studied separately, and in particular $\nu=-1/2$ is the fermionic zero mode and we have used Neumann boundary condition.

%We now study the special cases $\nu=\pm 1/2$. First for $\nu=1/2$,  two independent solutions can be chosen as 
%\bea
%u_{1/2}(\sigma) &=& (\zeta-1)^{1/2} , 
%\\
%v_{1/2}(\sigma)  &=& \frac{1}{2} (\zeta-1)^{-1/2} - \frac{(\zeta-1)^{1/2}}{4} \ln \left(
%\frac{\zeta+1}{\zeta-1}\right) . 
%\eea
%The Wronskian is  
%\be
%u_{1/2}(\sigma) v'_{1/2}(\sigma) - u'_{1/2}(\sigma) v_{1/2}(\sigma) = 1 . 
%\ee
%So the frequency is 
%\be
%\omega^{F1}_{1/2} = \ln \left[ u_{1/2}(\epsilon_0) v_{1/2}(L) \right] . 
%\ee
%Let us turn to $\nu=-1/2$. The solutions are 
%\bea
%u_{-1/2}(\sigma) &=& (\zeta+1)^{1/2}
%\\
%v_{-1/2}(\sigma)  &=& -\frac{1}{2} (\zeta+1)^{-1/2} + \frac{(\zeta+1)^{1/2}}{4} \ln \left(
%\frac{\zeta+1}{\zeta-1}\right) . 
%\eea
%Again the Wronskian is 
%\be
%u_{-1/2}(\sigma) v'_{-1/2}(\sigma) - u'_{-1/2}(\sigma) v_{-1/2}(\sigma) = 1 .
%\ee
%And the frequency is 
%\be
%\omega^{F1}_{-1/2} = \ln \left[ u_{-1/2}(\epsilon_0) v_{-1/2}(L) \right] . 
% \ee

For the other fermionic determinant $\omega^{F3}_\nu$, we may employ the following reparametrization 
\be
\psi(\sigma) = (\zeta+1)^{\nu/2-1/4}(\zeta-1)^{-\nu/2-1/4} y(\zeta) . 
\ee
%Then the corresponding homogeneous equation becomes 
%\be
%(\zeta^2-1) y'' + (\zeta-2\nu) y' = 0 .
%\ee
Again the differential equation is easily solved, and we choose the basis 
\bea
u_\nu (\sigma) &=& (\z+1)^{\nu/2-1/4} (\z-1)^{-\nu/2-1/4} , 
\\
v_\nu (\sigma) &=& (\z+1)^{\nu/2-1/4} (\z-1)^{-\nu/2-1/4}
\int^\zeta_{\zeta_0} \frac{(x-1)^{\nu-1/2}}{(x+1)^{\nu+1/2}} dx . 
\eea
%The Wronskin is independent of $\nu$ or the integration constant $\zeta_0$, and 
%\be
%u_\nu(\sigma) v'_\nu(\sigma) - u'_\nu(\sigma) v_\nu(\sigma) = -1 . 
%\ee
%And the solution with appropriate initial condition is 
%\be
%\psi(\sigma) = - u_\nu(\epsilon_0) v_\nu(\sigma) + v_\nu(\epsilon_0) u_\nu(\sigma) . 
%\ee
Except for $\nu=-1/2$ which is zero-mode, the frequency is then (before taking $L\ra \infty$ limit)
\be
\omega^{F3}_\nu = \ln \left[
e^{\nu(\epsilon_0+L)}\left( \sinh\epsilon_0\sinh L \right)^{1/2}
\int^{\coth\epsilon_0}_{\coth L} \frac{(x-1)^{\nu-1/2}}{(x+1)^{\nu+1/2}} dx
\right] . 
\label{f3}
\ee
And in the limit $L\ra\infty$, 
\be
\int^{\coth\epsilon_0}_{\coth L} \frac{(x-1)^{\nu-1/2}}{(x+1)^{\nu+1/2}} dx
=
\left\{
\begin{array}{lr}
B(e^{-2\epsilon_0};\nu+1/2,0)  & \nu\ge 1/2 ,
\cr
%-\ln (1-e^{-2\epsilon_0})   &  \nu=1/2 ,
%\cr
%2L & \nu=-1/2 ,
%\cr
-\frac{1}{\nu+1/2} e^{-2L(\nu+1/2)}  & \nu < -1/2 . 
\end{array}
\right . 
\ee
and one should substitute this into \eqref{f3}. Here we have expressed the integral 
in terms of the incomplete beta function,
\be
B(x;a,b) \equiv \int^x_0 t^{a-1} (1-t)^{b-1} dt . 
\ee
%One can for instance check that
%\bea
%\int^{\coth b}_{\coth a} \frac{(x-1)^{\nu-1/2}}{(x+1)^{\nu+1/2}} dx
%&=&
%B(1-e^{-2a};0,\nu+1/2) - B(1-e^{-2b};0,\nu+1/2)
%\nn\\
%%&=&
%B(e^{-2b};\nu+1/2,0) - B(e^{-2a};\nu+1/2,0) . 
%\eea
Since $\nu$ is half-integer for our purposes,  we may do 
the integration explicitly and obtain  \footnote{It is also a special case of the Lerch $\Phi$-transcendent, i.e. $ {}_2 F_{1} (n,1;n+1;x) = n \, \Phi(x,1,n) $. $\Phi$ is defined as 
$
\Phi ( z ,  s , a ) \equiv \sum_{k=0}^{\infty} \frac{z^k}{(a+k)^s} . 
$
} 
\be
B(x;n,0) =  \sum_{k=n}^\infty \frac{x^k}{k} = \left(\frac{x^n}{n}\right) \,{}_2 F_{1} (n,1;n+1;x)
. 
\ee

For $\nu=-1/2$ we study separately with  Neumann boundary condition. Summarizing, we have
 \be
 \omega^{F3}_{\nu} = \left\{
 \begin{array}{lr}
 \ln \left[ \frac{e^{(\nu+1/2)(L-\ep_0)}}{2(\nu+1/2)}
 \sqrt{\frac{2\ep}{1+\ep}} \cdot {}_2 F_{1} (n,1;n+1;\frac{1-\e}{1+\e})
 \right] , & \nu \ge 1/2
 \cr 
 \ln \left[ \sqrt{\frac{2\ep}{1+\ep}} \right],  & \nu=-1/2
 \cr
 \ln \left[ \frac{e^{-(\nu+1/2)(L-\ep_0)}}{-2(\nu+1/2)} \sqrt{\frac{2\ep}{1+\ep}}\right], & \nu < -1/2
 \end{array}
 \right.
 \ee
%These functions are special cases of the Lerch transcendent, which are defined as 

%and substitute this result into \eqref{f3}. 
%%%%%%%%%%%%%
\subsection{The regularized action}
%%%%%%%%%%%%%%%%%%%
We are now ready to go back to \eqref{reg} and evaluate the finite part. First of all,
one can easily convince oneself that 
\be
G' = 2(L-\ep_0) , 
\ee
so $G'$ should have no contribution after regularization. 

It turns out that $G_n\ra 0$ for large $n$, but the series $\sum G_n$ for given $\e$ is logarithmically divergent. We do the sum for large $\La$ and drop terms proportional to $\ln \La$. 
After rather tedious but straightforward computation, we may rewrite $\sum G_n$ in the following way. 
\bea
\sum_{n=0}^{\La} G_n  
&=& \frac{1}{2}\ln \left[ \frac{2^{7}(1+\ep) e^{-4L}}{ (1-\ep)^2(1+3\ep)}
\right]  + \frac{3}{2}(\La-1)\ln \left[\frac{\e}{1+\e}\right] + \frac{1}{2}\sum_{n=2}^{\La} S_n -\frac{3}{2} \sum_{n=0}^{\La} T_n ,
\label{act}
\\
S_n &=& \ln 
\left[
\frac{(n+\frac{1}{\epsilon})^4(n+1)(n-1)^4}{n^7 (n+\tfrac{1+\epsilon}{2\epsilon})
(n+\tfrac{1-\epsilon}{2\epsilon}) 
}
\right] ,
\\
T_n &=& \ln \left[\left(\frac{2\ep}{1+\ep}\right)^2
f_n(\ep) f_{n+1} (\ep) 
\right] .
\eea
Here we have introduced a shorthand notation $f_n(\e) = {}_2 F_{1} (n,1;n+1;\frac{1-\e}{1+\e})$. $S_n,T_n$ are chosen such that they converge to zero as $n\ra\infty$.
%\be
%f_n(\ep) = \left\{
%\begin{array}{ll}
%n \, \Phi( \tfrac{1-\ep}{1+\ep} , 1 , n )   & \, n>0 ,
%\cr
% 1  & \, n=0 . 
%\end{array}
%\right. 
%\ee
%Note that it is arranged so that $T_n\ra 0$ as $n\ra \infty$. 
%The sum is log-divergent, i.e. $\sum^\La T_n \sim f(\ep) \ln \La + g(\ep)$. 

One can see that the total sum is independent of cutoff $L$, as it should be the case. 
It is also obvious that the finite part of the first term in \eqref{act} is $\tfrac{7}{2}\ln 2$. The large-$\La$ behavior of $\sum S_n$ can be studied using Stirling's formula. After we drop the terms proportional to $\ln \La, 1/\ep, \ln \ep$ etc,
\bea
\frac{1}{2} \left(\sum_{n=2}^{\La} S_n \right)_{reg} 
%&=& 0 + 0 + \frac{1}{4} \ln \frac{\pi}{2} + \ln 2\pi - \frac{7}{4} \ln 2\pi - \ln 2 - \frac{1}{2} \ln 2 
%\nn\\
&=& -\frac{1}{2} \ln (32 \pi) .
\label{analytic}
\eea

%%%%%%%%%%%%%%%%%%%%%%%%%%%%%%%%%
\begin{figure}
\centering
\includegraphics[scale=1.3,bb= 0 0 190 126, clip=true]{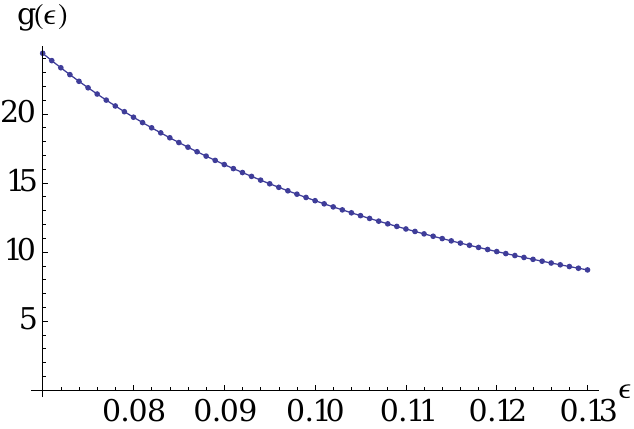}
\caption{Dots represent $g(\e)$ from least-square fit of numerical sum $\sum^\Lambda T_n$ against $f(\e)\ln \Lambda + g(\e)$. We used values $\Lambda=2500,2510,\cdots, 3500$. Solid line corresponds to  $g(\e)=0.8182 - 1.039 \frac{1}{\e} - 1.011 \frac{\ln \e}{\e}$. We calculated for 61 points in $0.07\le \e \le 0.13$ and find $\chi^2=1.68\times 10^{-9}$.
%in the range of $0.06<\e<0.13$
}
\label{logx}
\end{figure}
%%%%%%%%%%%%%%%%%%%%%%%%%%%%%%%%%
For the summation of $T_n$, unfortunately we are not able to find the sum in closed form for large $\La$. We will resort to numerical methods. Our strategy is as follows. We first fix $\e$ and consider $\Lambda\ra\infty$. Since the sum is log-divergent, 
\be
\sum_{n=1}^{\Lambda} T_n = f(\e) \ln \Lambda +g(\e) + (\mbox{subleading in } {\Lambda}) . 
\ee
We can read off $f(\e),g(\e)$ from a least-square fit after evaluating the sum numerically for a number of large values for $\Lambda$. To obtain a  regularized value, we now concentrate on $g(\e)$. This function is also divergent as $\e\ra\infty$, and we find it is very closely approximated by $g(\e)= \alpha + \beta \frac{\ln \e}{\e}+ \gamma \frac{1}{\e} + \delta \log \frac{1}{\e}$ in the leading orders. The numerical results versus this curve is shown in Figure \ref{logx}.
When we implement this method however, one has to be careful since the result depends rather sensitively on the choice of cutoffs $\Lambda$ and $\e$. It is not surprising since the series is not convergent,  after all. We want to send $\e\ra 0$ eventually, but since we take $\Lambda\ra\infty$ first, $\e$ should not be too small, i.e. $\e\Lambda\gg 1$ should be always satisfied. We 
have tried different ranges for $\Lambda,\e$ until the result for $\tfrac{1}{2}\sum S_n$ is reasonably close to the analytic result. For $2500\le\Lambda\le 3500$ and $0.07\le\e\le 0.13$, our numerical result is $-2.2432$, where the exact value is $-\tfrac{1}{2}\ln 32\pi= -2.3052$. We use the same values of $\Lambda,\e$ to evaluate $\sum T_n$, and obtain the final result
\be
\G_{reg} =  -1.106
\label{result}
\ee 
%%%%%
\section{Discussion}
\label{discuss}
%%%%%%%%%%%%%%
$\Gamma_{reg}$ should be 
compared to the field theory result $\ln 2 = 0.6931$, and certainly the difference is not negligible. Recall that for 1/2-BPS Wilson loop in $AdS_5\times S^5$, there was also a discrepancy %of $\ln 2$  
and it was deemed to come from the normalization of the zero modes \cite{Kruczenski:2008zk,Drukker:2000rr,Zarembo:2002an}. As far as we know, this coefficient is not determined for $AdS_4\times \mathbb{CP}^3$, let alone $AdS_5\times S^5$ in Type IIB.
We note that the normalization convention of holographic Wilson loops in ABJM theory was discussed in Section 5.2
of \cite{Drukker:2010nc}. 

To bypass the normalization problem and check the validity of string one-loop computations, we may study other supersymmetric Wilson loop operators and calculate the ratio between physically different BPS Wilson loops.  In the field theory description, there are 1/6-BPS Wilson loops with $\langle W\rangle \approx \sqrt{\lambda/2}\exp(\sqrt{2\pi^2 \lambda})$. While 1/2-BPS Wilson loops are pointlike in $\mathbb{CP}^3$ and break the global symmetry $SU(4)$ into $SU(3)$, 1/6-BPS ones preserve only $SU(2)$ and it is natural to expect that they are smeared over $\mathbb{CP}^1\in\mathbb{CP}^3$ \cite{Drukker:2008zx,Rey:2008bh}. We plan to construct such classical string solutions in $AdS_4\times \mathbb{CP}^3$ explicitly and study its fluctuations in a separate publication. 

Although we only studied circular Wilson loops in detail here, \eqref{pf1} is valid for a straight line as well and our computation is easily extendable to the $D=3$ analog of quark-antiquark potential calculation using holography. Of course in principle a similar computation in string theory side can be checked against the localization calculation for any supersymmetric Wilson loops. Let us emphasize that recently 
a two-parameter family of string solutions interpolating the circle and a pair of straight line Wilson loops in $N=4, D=4$ super Yang-Mills theory was studied in \cite{Drukker:2011za,Forini:2012rh}. It is also pointed out that the angle dependence of general BPS Wilson loop operators can be related to interesting physical quantities such as cusp anomalous dimension, radiation emitted by a moving quark etc. \cite{Correa:2012at,Fiol:2012sg,Correa:2012nk,Drukker:2012de,Correa:2012hh}. For a recent study of cusp anomalous dimension in ABJM model, see \cite{Forini:2012bb}. With such  applications in mind, it will be intriguing to construct general BPS Wilson loops and pursue their exact evaluation.

%%%%%%%%%%%%%%%%%%%%%%%%%%%%%%%%%
%\section{Discussion}
%\label{discussion}
%%%%%%%%%%%%%%%%%%%%%%%%%%%%%%%%%

\begin{acknowledgments}
%The research of HK was supported by a post-doctoral fellowship grant from Kyung Hee University (KHU-20110694).
The research of NK and JHL is supported by the National Research Foundation of Korea (NRF) funded by the Korean Government (MEST) with grant No. 2009-0085995, 2010-0023121, and also through the Center for Quantum Spacetime (CQUeST) of Sogang University with grant No. 2005-0049409. %NK also gratefully acknowledges the hospitality of the Institute for Advanced Study, where part of this work was completed. 
\end{acknowledgments}

%%%%%%%%%%%%%%

%%%%%%%%%%%%%%%%

%%%%%%%%%%%%%%%%%%%%%%%%%%%%%%%%%%%%%%%%%%%%%%%%%%%%%%%%%%%%%%%%%%%%%%%%%%%%%%%%%%%%
%%%%%%%%%%%%%%%%%%%%%%%%%%%%%%%%%%%%%%%%%%%%%%%%%%%%%%%%%%%%%%%%%%%%%%%%%%%%%%%%%%%%

%%%%%%%%%%%%%%%%%%%%%%%%%%%%%%%%%%%%%%%%%%%%%%%%%%%%%%%%%%%%%%%%%%%%%%%%%%%%%%%%%%%%%%%%%

%%%%%%%%%%%%%%%%%%%%%%%%%%%%%%%%%%%%%%%
%%%%%%%%%%%%%%%%%%%%%%%%%%%%%%%%%%%%%%%

%%%%%%%%%%%%%%%%%%%%%%%%%%%%%%%%%%%%%%%

%%%%%%%%%%%%%%%%%%%%%%%%%%%%%%%%%%%%%%%%%%%%%%%%%%

%%%%%%%%%%%%%%%%%%%%%%%%%%%%%%%%%%%%%%%%%%%%%%%%%%%%%%%%%%%%%%%%%
\bibliographystyle{utphys}
\bibliography{wl}{}
%%%%%%%%%%%%%%%%%%%%%%%%%%%%%%%%%%%%%%%%%%%%%%%%%%%%%%%%%%%%%%%%%
\end{document}